\documentclass[useAMS,usenatbib]{mn2e}

\usepackage[utf8x]{inputenc}
\usepackage{bm}
\usepackage{latexsym}
\usepackage{tabularx}
\usepackage{dcolumn}
\usepackage{amsmath,amsfonts,amssymb}
\usepackage{graphicx,epsfig}
\usepackage{color}
\usepackage{subfigure}
\usepackage{sidecap}
\usepackage{longtable}
\usepackage{natbib}
\usepackage{float}

\newcommand{\de}{{\rm d}}
\newcommand{\lya}{{Lyman-$\alpha$}}
\newcommand{\bea}{\begin{eqnarray}}
\newcommand{\eea}{\end{eqnarray}}
\newcommand{\be}{\begin{equation}}
\newcommand{\ee}{\end{equation}}
\newcommand{\f}{\frac}
\newcommand{\df}{\dfrac}

\newcommand{\bc}{\begin{center}}
\newcommand{\ec}{\end{center}}

\newcommand{\T}{\rule{0pt}{3.6ex}}

\newcommand{\B}{\rule[-1.0ex]{0pt}{0pt}}

\title[Clustering of Lyman-$\alpha$ emitters]{Clustering at high redshift: The connection between Lyman Alpha emitters and Lyman break galaxies}
\begin{document}

\author[C. Jose et al.]
{Charles Jose$^1$\thanks{charles@iucaa.ernet.in},
Raghunathan Srianand$^1$\thanks{anand@iucaa.ernet.in} 
and Kandaswamy Subramanian$^1$\thanks{kandu@iucaa.ernet.in}\\ 
%and \newauthor Saumyadip Samui$^2$\thanks{samuis@ukzn.ac.za} \\
$^1$IUCAA,Post Bag 4, Pune University Campus, Ganeshkhind, Pune 411007, India\\
%$^2$Astrophysics and Cosmology Research Unit, School of Physics, UKZN, Durban 4001, 
%South Africa}
}
\maketitle

\begin{abstract}
We present a physically motivated semi-analytic model to understand the 
clustering of high redshift Lyman-$\alpha$ Emitters (LAEs). We show that the 
model parameters constrained by the observed luminosity functions, can be 
used to predict large scale bias and angular correlation function of LAEs. 
These predictions are shown to reproduce the observations remarkably well. 
We find that average masses of dark matter halos hosting LAEs brighter than   
threshold narrow band magnitude $\sim 25$ are $\sim 10^{11}M_\odot$. These are  
smaller than that of typical Lyman Break Galaxies (LBGs) brighter than similar 
threshold continuum magnitude by a factor $\sim 10$.  
This results in a smaller clustering strength of LAEs compared to LBGs. 
However, using the observed relationship between UV continuum and Lyman-$\alpha$ 
luminosity of LAEs, we show that both LAEs and LBGs belong to 
the same parent galaxy population with narrow band techniques having greater 
efficiency in picking up galaxies with low UV luminosity. 
We also show that the lack of evidence for the presence of the one halo 
term in the observed LAE angular correlation functions can be attributed 
to sub-Poisson distribution of LAEs in dark matter halos as a result of 
their low halo occupations.
\end{abstract}


\section{Introduction}

Over the past decade there has been a growing wealth of observations 
probing the properties of high redshift galaxies 
\citep{steidel_LBG_99,giavalisco_GOODS_04,ouchi_LBG_04,beckwith_hudf,
grogin_CANDELS, illingworth_hxdf}. Various surveys, 
using the Lyman break color selection technique 
\citep[see for example,][]{madau_96, steidel_96_1,
steidel_98,steidel_98_1}, provided fairly good estimates of 
luminosity functions (LF) up to $z\sim 8$ 
\citep{bouwens_07_LF_z46,bouwens_07_LF_z710,reddy_08_LF,
mclure_hudf12_12,schenker_uvlfz78_hudf12_13,
oesch_hudf12_13,lorenzoni_uvlfz79_candels_13} and spatial clustering 
up to $z\sim 5$ \citep{giavalisco_dickinson_01,ouchi_04_acf,
ouchi_hamana_05_acf,kashikawa_06_acf,hildebrandt_09_acf,
savoy_11_acf,bielby_11_acf} of these Lyman break galaxies (LBG). 
On the other hand, narrow-band searches 
for high redshift \lya\ line \citep{hu_lae_1998, rhoads_lae_00,ouchi_lae_03, 
shimasaku_lae_06,gronwall_lae_07,finkelstein_lae_07,dawson_lae_07,  
ouchi_lae_08,shioya_lae_09,wang_lae_09, hu_lae_10, ouchi_lae_10} 
have detected substantial number of \lya\ emitters (LAEs). 
These surveys helped to infer the statistical properties of LAEs, 
particularly their UV and \lya\ luminosity functions 
and angular correlation functions \citep{ouchi_lae_03, ouchi_lae_08, 
shioya_lae_09,wang_lae_09,ouchi_lae_10}. While LBGs selected through
colour cuts are biased towards bright galaxies, the narrow band
selection of LAEs  is biased towards galaxies having strong 
Lyman-$\alpha$ emission
line and weak continuum emission. Therefore, these two techniques
pickup galaxies with different type of selection biases. 
Despite several detailed studies there is no clear consensus on
whether there is any differences between the two populations based,
%KS: added comma
on properties such as stellar mass, dust content, age and 
star formation rate etc \citep[see for example,][]{Gawiser06,Pentericci07,
ouchi_lae_08,Kornei10}.

In the hierarchical model of structure formation 
the statistical properties of galaxies are
determined by that of the parent dark matter halo population, given a 
prescription  for how stars form inside these halos. 
The properties of dark matter halos are quite well understood 
using N-body simulations and analytical models like the halo 
model of large scale structure. There have been a number of studies 
over the past years, probing the properties of LAEs using semi-analytical 
and numerical (simulation) methods \citep{shimasaku_laez5.7_06,
delliou_lae_06,mao_lae_07,
kobayashi_lae_07,samui_07,dayal_lae_08,tilvi_lae_09,samui_09_lae,
zheng_lae_2010,hu_lae_10, nagamine_lae_10,garel_lae_10, 
romero_fesclya_2011, kobayashi_lae_12,dijkstra_lae_12}. 
These models have been successful in reproducing the (i) UV 
LF of LBGs, (ii) UV LF of LAEs and (iii) Lyman-$\alpha$ LF of LAEs. 
We also have been extensively developing simple physical models of galaxy 
formation to understand the LFs of LBGs and Lyman-$\alpha$ emitters. 
Any galaxy formation model that can simultaneously  explain the observed 
properties \citep[like for example,][]{Dayal12} of LBGs and LAEs is very 
useful to further our understanding of physics of galaxy formation 
at high-$z$.
%These approaches provide the abundance, 
%spatial distribution and merger history of dark matter halos. 
%
%Numerical simulations also suggest a possible universal dark matter 
%halo density profile, NFW profile \citep{NFW_97}. Given the above inputs on
%dark matter halo properties and a specific model of galaxy formation 
%inside these halos, it is possible to explain the two major observables 
%of galaxies, their luminosity function and clustering. 
%%In addition, 
%such models can throw light on the complex physics of galaxy formation, 
%such as rate and duration of star formation, feedback mechanisms etc. 

In particular by modelling the clustering of LBGs and LAEs one will
be able to understand any differences in the range of halo masses probed
by these two population of sources. 
Recently we have presented a physically motivated semi-analytic
halo model to simultaneously explain the LF and spatial clustering
of high-$z$ LBGs \citep[][hereafter J13]{charles_13_LBG}. 
Here we extend this model
to explain the measured LFs and angular correlation function
of high-$z$ LAEs.
The organization of this paper is as follows. In the next section we 
briefly outline
our models for computation of the LF of LAE. In Section 3 we 
extend our models for LBG clustering to study the clustering of LAEs.
Section 4 presents a comprehensive comparison of the total angular correlation 
function computed in various models with observations. A discussion of our 
results and conclusions are presented in the final section. 
For all calculations we adopt a flat $\Lambda$CDM universe 
with cosmological parameters consistent with 7 year Wilkinson 
Microwave Anisotropy Probe (WMAP7) observations \citep{larson_11_wmap7}. 
Accordingly we assume $\Omega_m=0.27$, $\Omega_\Lambda=0.73$, 
$\Omega_b=0.045$, $h=0.71$, $n_s = 0.963$ and $\sigma_8 = 0.801 h^{-1}$Mpc. 
Here $\Omega_i$ is the background density of any species 
'i' in units of critical density $\rho_{c}$. 
The Hubble constant is $H_0 = 100 h$ km s$^{-1}$ Mpc$^{-1}$

\section{The Star formation rate and Luminosity function} 
\label{sec:lf}

Here we recall the salient points of our semi-analytic model 
used to explain the observed luminosity function of high-$z$ 
LBGs \citep[][hereafter SSS07]{samui_07} and LAEs 
\citep[][hereafter SSS09]{samui_09_lae}.
%
%Our aim is to construct a self consistent semi-analytical model, 
%that can explain the luminosity functions and clustering of high 
%redshift LAEs. In this section, we briefly recall the semi analytical 
%treatment of \citet{samui_09_lae} (hereafter SSS09) to model 
%Ly$\alpha$ LFs of high redshift LAEs. 
%Firstly, we describe 
%the models of \cite{samui_07} (hereafter SSS07) for UV LF of 
%high-z LBGs, as they are crucial for computing the corresponding 
%Ly$\alpha$ LFs. In the models of SSS07 and SSS09, 
The star formation 
rate ($\dot M_{SF}$) in a dark matter halo of mass $M$ collapsed at 
redshift $z_c$ and observed at redshift $z$ is given by 
\citep[see,][]{chiu_00,choudhury_02},
\bea
\dot M_{SF}(M,z,z_c) &=& f_\ast \left(\frac{\Omega_b}{\Omega_m} M \right) 
                \frac{t(z)-t(z_c)}{\kappa^2 t^2_{dyn(z_c)}} \label{eqn:SFR}
 \\ \nonumber
  && \times \exp\left[-\frac{t(z)-t(z_c)}
                {\kappa t_{dyn(z_c)}}\right], 
\eea
where, the amount and duration of the star formation is determined
by values of $f_*$ and $\kappa$ respectively. 
%As in SSS07 we fix these 
%two parameters by fitting the observed UV luminosity functions of LBGs. 
Further, $t(z)$ is the age of the universe; thus $t(z)-t(z_c)$ gives
the age of the galaxy at $z$ that was formed at an earlier epoch $z_c$,
and $t_{dyn}$ is the dynamical time at that epoch.
The stars are formed with a Salpeter initial mass function (IMF) having  
the mass range $1 - 100 ~M_\odot$. The star formation 
rate of a galaxy, as given in Eq.~(\ref{eqn:SFR}), is then
%convolved 
used
%with burst luminosity of unit solar mass 
to obtain the UV luminosity at 1500\AA\ (i.e $L_{1500}$) and AB magnitude
($M_{AB}$) at any given time using the procedure described in detail in SSS07. 
%We use the parameter $\eta$ to take into account the dust correction.
%
%The population synthesis code {\sc Starburst99} \citep{starburst_99} is used 
%to obtain the rest frame UV luminosity ($l_{1500}$) at 1500 \AA\ as a function 
%of time of a galaxy undergoing a burst n of star formation. The star formation 
%rate of a galaxy, as given in Eq.~(\ref{eq1}), is then convolved 
%with this burst luminosity to get the time evolution of the luminosity, 
%$L_{1500}$, of an individual star forming galaxy (See Eq.~(6) and Figure 1 of SSS07)
The observed luminosity is only a fraction $1/\eta$ of the actual luminosity
because of the dust reddening in the galaxy. 
%This luminosity ($L = L_{1500}/\eta$) is then converted to a standard 
%absolute AB magnitude $M_{AB}$,  using the equation given by \citep{oke_83},
%to enable direct comparison with the observed data. 
Having obtained the $M_{AB}$ of individual galaxies we can compute
the luminosity function $\Phi(M_{AB}, z )$ at any redshift $z$ 
using,
\bea
\Phi(M_{AB}, z ) dM_{AB}&=& \int\limits_z^\infty dz_c  \frac{dn(M(M_{AB}),z_c)}{dz_c}
\frac{dM}{dL_{1500}} \\ \nonumber  &&\times~ \frac{dL_{1500}}{dM_{AB}} ~dM_{AB}.
\eea
Here $\de n(M,z_c)/\de z_c =  \dot n(M,z_c) dt/dz_c$, and $\dot n(M,z_c) dM$ is 
the formation rate of halos in the mass range $(M, M+dM)$ at redshift $z_c$. 
%SSS07 modelled 
We model this formation rate as the time derivative of 
\citet{sheth_tormen_99} (hereafter ST) mass function as they are found to be 
good in reproducing the observed LF of high-$z$ LBGs. 
Therefore we use $\dot n(M,z_c) = \de n_{ST}(M,z_c)/\de t$  where 
$ n_{ST}(M,z_c)$ is the ST mass function at $z_c$. 
Also note that we use the notation $n(M)$ for $dn/dM$ for convenience. 

Star formation in a given halo also depends on the cooling efficiency of 
the gas and various other feedback processes. We assume that gas in halos 
with virial temperatures ($T_{vir}$) in excess of $10^4$ K can cool 
%(due to recombination line cooling from hydrogen and helium) 
and collapse to form stars.
% However the ionization of the IGM by UV photons 
%increases the temperature of the gas thereby increasing the Jean's mass 
%for collapse. Thus in ionized regions, we incorporate this feedback by
In the ionized regions we incorporate radiative feedback by 
a complete suppression of galaxy formation in halos with circular velocity 
$v_c \leqslant 35$ km s$^{-1}$ and no suppression with $v_c  \geqslant V_u 
= 95$ km s$^{-1}$ \citep{bromm_02}. For intermediate circular velocities, 
a linear fit from $1$ to $0$ is adopted as the suppression factor 
[\cite{bromm_02}; see also \cite{benson_02,dijkstra_04}, SSS07]. 
%SSS07 found that this feedback mechanism naturally leads to the 
%observed flattening of the LF at the low luminosity end.
%In our models, 
In addition, we also incorporate the possible Active Galactic Nuclei (AGN) 
feedback that suppresses star formation in the high mass halos, by multiplying 
the star formation rate by a factor $[1+M/M_{agn}]^{-0.5}$ (see J13).  
This decreases the star formation activity in 
high mass halos above a characteristic mass scale $M_{agn}$, which is believed 
to be $\sim 10^{12} M_\odot$ \citep[see][]{bower_06,best_06}. 

The crucial parameter of our model is $f_\ast/\eta$ which governs the mass 
to light ratio of galaxies at any given redshift. 
Recently, some tentative evidences for dust obscuration in LBGS as 
a function of UV luminosity and redshift are reported 
\citep{reddy_steidal_09,reddy_pettini_12,bouwens_illingworth_12}. 
%However, introducing this trend, which is not yet well established would 
%lead to additional uncertainity in our models. 
%Therefore, for simplicity, we fix $\eta$ to be luminosity independent. 
However, as introducing this not so well established trend would 
lead to additional uncertainity in our models, 
we fix $\eta$ to be luminosity independent. 
J13 showed that for a luminosity 
independent $\eta$, only the combination $f_\ast/\eta$ determines 
the clustering predictions of LBGs. 
This ($f_\ast/\eta$) is fixed by fitting the observed UV LF 
of LBGs using $\chi^2$ minimization.

% Note that for a luminosity independent $\eta$, 
%we only need the combined parameter $f_\ast/\eta$ to fit the LFs of LBGs. 
%In this way our physically motivated model for star formation gives the 
%relationship between the halo mass (M) and the luminosity (L) of the 
%galaxy it hosts. 

\begin{figure*}
\includegraphics[trim=0cm 0.0cm 0cm 0.0cm, clip=true, width =16cm, height=13cm, angle=0]
{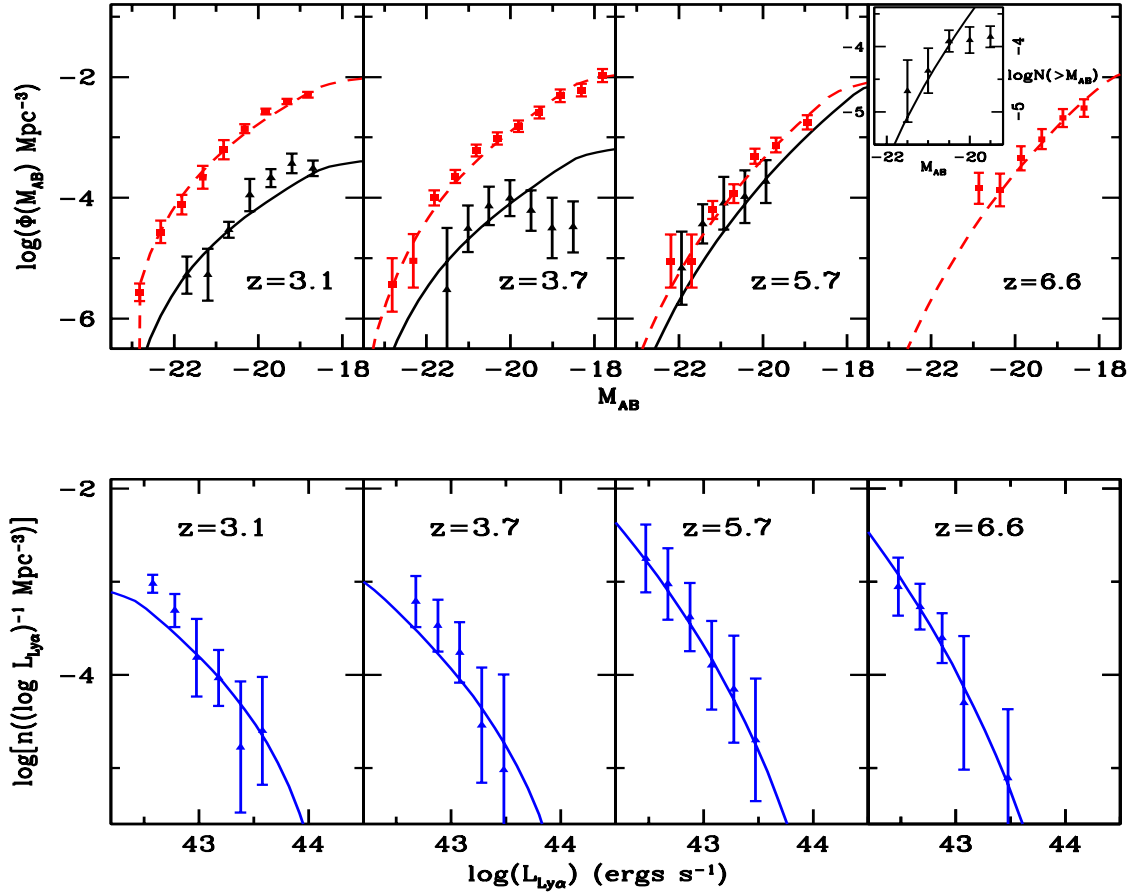}
\caption{
Top panels show the predicted UV LF of LBGs (dashed lines) at $z=3,4,6$ and $7$ 
and that of LAEs (thick lines) at $z=3.1,3.8,5.7$ and $6.6$ along with the observed 
data . The observed data points of LBGs taken from \protect \cite{bouwens_07_LF_z46,
bouwens_07_LF_z710} and \protect \cite{reddy_08_LF} are shown as red filled squares. 
The black filled triangles are the observed UV LF data points from 
\protect \cite{ouchi_lae_10,ouchi_lae_08}. 
The insert in the top right panel shows the cumulative UV LF of LAEs 
(solid black lines) at $z=6.5$ as predicted by our models along with the 
data given by \protect \cite{kashikawa_lae_lf_2011} (black filled triangles).  
Bottom panels show our model predictions of Lyman $\alpha$ LF of 
LAEs at $z=3.1,3.7,5.7$ and $6.6$. Here, the observed data points 
and error bars are from \protect \cite{ouchi_lae_10,ouchi_lae_08} 
}
\label{fig:lf}
\end{figure*}

%In order to compute the 
%We assume the case-B recombination to compute
%the Lyman-$\alpha$ luminosity of a star forming galaxy.
%we assume case-B recombination, 
%where two Lyman-$\alpha$ photons 
%are produced out of three hydrogen ionizing photons (Osterbrock, 1989) 
%that are confined within the interstellar medium of the galaxy. 
The Lyman-$\alpha$ luminosity produced in any star forming region 
(when we use the case-B recombination) is 
related to its instantaneous star formation rate and is given by, 
\begin{equation}
L_{Ly\alpha} = 0.67 h \nu_{\alpha} ( 1 - f_{esc} ) {N}_\gamma {\dot M}_{SF}.
\label{eqn_lyman}
\end{equation}
Here, $h \nu_{\alpha} = 10.2$~eV is the energy of a Lyman-$\alpha$ photon 
and we fix $f_{esc} = 0.1$, the escape fraction of UV ionizing photons.
Further, ${N}_\gamma$ is the production rate of ionizing photon
per unit solar mass of star formation which is a function of  
IMF and the metallicity. We use $N_\gamma = 10450 $ for the Salpeter IMF 
and metallicity of $Z = 0.0004$ (See SSS07).

The observed Lyman-$\alpha$ luminosity is given by, 
\begin{equation}
L_{Ly\alpha}^{obs} = f_{esc}^{Ly\alpha} L_{Ly\alpha}.
\end{equation}
Here, $f_{esc}^{Ly\alpha}$ is the escape fraction of the Lyman-$\alpha$ photons 
which is decided by the dust optical depth, velocity field of the ISM in the 
galaxies and the Lyman-$\alpha$ optical depth due to ambient intergalactic 
medium around the galaxies.

Observations suggest that only a small fraction of LBGs ($G_f$) show 
detectable Lyman-$\alpha$ emission 
\citep[See for example][]{kornei_10_lae}. 
%Therefore as in SSS09, we assume 
%that only a 
%fraction $G_f$ of the entire LBGs will be detected as LAEs in surveys. 
In our models we assume that the UV LF of LAEs at any $z$ is a 
fraction $G_f$ of the UV LF of LBGs at that redshift 
(i.e. $\Phi^{UV}_{LAE} = G_f \Phi^{UV}_{LBG}$). We estimate $G_f$ by 
comparing the model predictions of UV LF of LAEs with the observed 
data.  

As a fiducial model we have chosen $\kappa = 1.0$ at all redshifts and 
for $M_{agn}$, $0.8\times 10^{12}M_\odot$, $1.5\times 10^{12}M_\odot$ at 
redshifts $3$ and $4$. For higher redshifts we 
do not require any AGN feedback. As in SSS07 and J13, we 
find that with these parameters the observed luminosity functions 
is well reproduced for $f_\ast/\eta$ of 0.042, 0.038, 0.037 
and 0.044 respectively for $z=3,4, 6$ and $7$. We note that the values 
of  $f_\ast/\eta$ is nearly constant over the redshift range considered. 
These values are tabulated in  Table~\ref{tab1}. 
If we use the average value of $\eta$ at $z=3-7$ estimated from 
\cite{gonzalez_bouwens_12,reddy_pettini_12}, we get $f_\ast =0.13 - 0.05$ 
in the same redshift range. This indicates an increase in the baryon fraction
that is being converted into stars with time.
In the top panel of  Fig.~\ref{fig:lf} we show our model predictions of best 
fit UV LF of LBGs in dashed lines. The corresponding  observational data is 
shown as solid squares. 

The best fit values of $G_f$ that fits the UV LF of LAEs are also tabulated 
in Table~\ref{tab1}. Our model predictions of UV LF of 
LAEs are shown in the top panels (in solid black lines) of Fig.~\ref{fig:lf} 
along with the observational data.  
At $z=6.6$, we compare our model predictions of 
cumulative UV LF of LAEs with the observed 
cumulative UV LF of LBGs given by \cite{kashikawa_lae_lf_2011} and fix 
$G_f$ to be 0.5\footnote{Note that for predicting UV LF of LAEs at a given redshift 
we use the $f_*/\eta$ constrained by the observed UV LF of LBGs at a 
slightly different redshift. For example to predict the UV LF of LAEs at 
$z=3.1$ we use the $f_*/\eta$ obtained using $z=3$ observed UV LF of LBGs.} (See 
insert in the top right panel of Fig.~\ref{fig:lf}).   
From the table it is clear that, $G_f$, the fraction of LBGs visible as 
LAEs increases from $\sim$ $0.05$ to $0.5$ from redshift $3-7$. 
Such a trend has been noted in the previous works of 
\cite{stark_lae_gf_10, pentericci_lae_11, curtis-lake_lae_gf_12, 
romero_Gf_12}. 
These authors treat the fraction of LBGs visible as 
LAEs as a function of the UV magnitude of LBGs. However, the 
average fraction of fraction of LBGs visible as LAEs, obtained from 
their studies are consistent with our results.

Given $G_f$ one can obtain the Lyman-$\alpha$ LF of LAEs by varying 
$f_* f_{\rm esc}^{Ly\alpha}$ as a free parameter. In this way, the observed 
$Ly\alpha$ LFs of LAEs are well reproduced for 
$f_\ast f_{\rm esc}^{Ly\alpha}$ = $0.066,0.040,0.031$ and $0.029$ 
%KS: corrected the f*/fesc to f*fesc above
respectively at $z=3.1,3.8, 5.7$ and $6.6$. 
These best fit $f_* f_{\rm esc}^{Ly\alpha}$ are tabulated in Table \ref{tab1}.  
We also show in the lower panels of  Fig.\ref{fig:lf}, our models 
predictions of the best fit Ly$\alpha$ LFs of LAEs along with observed data 
at different $z$. 
Using the $f_\ast$ determined earlier we have obtained 
$f_{\rm esc}^{Ly\alpha} = 0.5 -0.6$ in the redshift range $z=3.1-6.6$. 
Thus our models predicts no strong evolution in 
$f_{\rm esc}^{Ly\alpha}$ from $z=3.1$ to $6.6$ which is consistent with the 
earlier studies of \cite{hayes_12_fesclae,blanc_2011_fesclae, ono_2010_fesclae},  
who also found no clear evolution of $f_{\rm esc}^{Ly\alpha}$ in the above 
redshift range.

\section{Two point correlation functions of LAEs}
\label{sec:xir}

%In this section we couple our models of Lyman $\alpha$ LF with the halo model 
%to compute the correlation function of high-$z$ LAEs. We have earlier shown that, 
%our simple model model for star formation in galaxies merged with the halo model, 
%can also explain the clustering of the high $z$ LBGs (\cite{charles_13_LBG}; 
%hereafter J13). Here we use the similar formalism to predict the angualr 
%correlation functons of LAEs.

In this section we compute the angular correlation functions 
of high-$z$ LAEs on all angular scales. To do this, one requires the full 
knowledge of the halo occupation distribution (HOD). The HOD describes the 
probability distribution for galaxies brighter than a given 
luminosity threshold to reside in a dark matter halo of 
mass $M$ \citep{bullock_02,berlind_02}. 
Given the HOD, one can compute the average number of 
galaxies brighter than the luminosity threshold hosted by a dark matter halo 
of mass $M$. Further this information can be used with the halo model to 
calculate the galaxy correlation functions.  

%In order to compute the galaxy-galaxy correlation function on all 
%scales one requires the full knowledge of the halo occupation distribution 
%(HOD), which describes the conditional probability $P(N|M)$ for $N$ galaxies 
%of a given type to reside inside a halo of mass $M$ \citep{bullock_02,berlind_02}. 

We modify the physically motivated HOD models of LBGs 
J13 to compute the same for LAEs that are 
brighter than a threshold Luminosity $L^{Ly\alpha}_{th}$.
%
%
%
%
%The clustering data provided by \cite{ouchi_lae_10} is for LAEs above a 
%luminosity threshold. Therefore, our first aim is to compute the HOD for LAEs 
%above a given luminosity threshold $L^{Ly\alpha}_{th}$ (The threshold luminosity  
%corresponds to threshold magnitude). In J13, we have provided 
%a physically motivated model to compute the HOD for LBGs above a given luminosity 
%threshold. We briefly recall this approach in the context of 
%Lyman $\alpha$ emitting galaxies. 
%
%In the previous sections we have assumed that a halo can host at most one 
%detectable galaxy. This assumption is not adequate to explain clustering 
%at small angular scales (see J13), especially on scales smaller than 
%virial radius. Each halo can in principle host multiple galaxies. 
%Following the approachs of ~\citet{kravtsov_04,zheng_05,cooray_ouchi_06},
%we separate the central and satellite contributions to HOD (see also J13). 
%That is 
%
The mean number of galaxies inside a dark matter halo can be written 
as $N_g(M) = f_{cen}(M)+N_{s}(M)$ where $f_{cen}(M)$ and $N_s(M)$ are respectively 
the mean number of central and satellite galaxies. We also assume the central 
galaxy to be situated at the center of the halo and satellite galaxies following 
NFW density distribution of dark matter matter halos \citep{kravtsov_04}. 

\subsection{The central galaxy occupation}

\begin{figure}
\includegraphics[trim=0cm 0cm 0cm 0cm, clip=true, width =8.0cm, height=7.5cm, angle=0]
{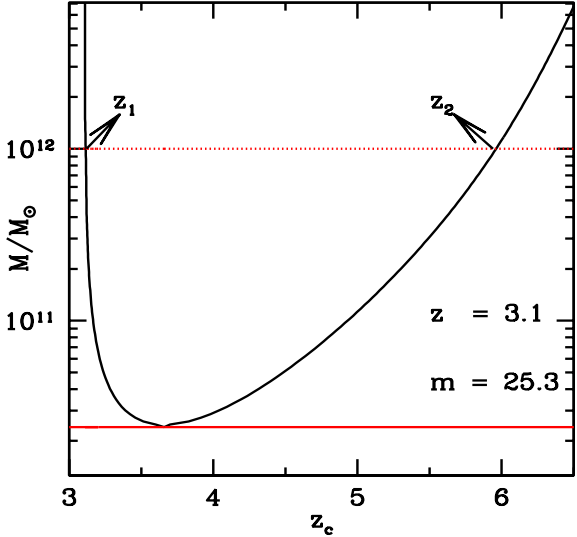} 
\caption{Masses of the dark matter halos that shine with an apparent magnitude 
$m^{Ly\alpha}_{AB}=25.3$ at $z=3.1$ as a function of their formation (collapse) 
redshift. Halos of a particular mass formed at two different redshifts can 
shine with the same brightness at $z=3.1$ which we refer as  $z_1$ and $z_2$. 
For example, when $M=10^{12} M_\odot$ (shown by dotted red horizontal line)
these two redshifts are approximately 3.13 and 5.94. The figure also shows 
in red horizontal solid line the minimum mass of the galaxy that can produce  
a magnitude 25.3 at $z=3.1$ which is roughly $2.4 \times 10^{10}  M_\odot$.
}
\label{fig:zc_mass}
\end{figure}

In our model the luminosity $L$ of a galaxy initially increases and 
then decreases with its age. Therefore, a galaxy of mass $M$ collapsed 
at two different redshifts can shine with the same luminosity $L$ at a 
given observational redshift. Depending on the redshift of 
formation these galaxies will be either in their increasing phase or 
decreasing phase of luminosity. In between these two epochs (i.e. two redshifts) 
the galaxy will shine with luminosity greater than $L$. For such a model, 
the mean occupation number of central galaxies with 
a Lyman-$\alpha$ luminosity above a threshold luminosity 
(or magnitude) $L^{Ly\alpha}_{th}$ inside a halo of mass $M$ at any 
redshift $z$ as in Eq.~(15) of J13 is
\be
f_{cen}(L^{Ly\alpha}_{th},M,z) = G_f \f{\int\limits_{z_1}^{z_2} dz_c \df{dn(M,z_c)}{dz_c}} 
              {n(M,z)}.       \label{eqn:fcen}
\ee
Here $z_1(M,L^{Ly\alpha}_{th},z)$ and $z_2(M,L^{Ly\alpha}_{th},z)$ are the two 
redshifts at which a galaxy of mass $M$ is to be formed, so that it shines with 
an observed luminosity $L^{Ly\alpha}_{th}$ at $z$. Note a galaxy of mass $M$ 
that has collapsed between these two redshifts $z_1$ and $z_2$ will shine with 
a luminosity greater than $L^{Ly\alpha}_{th}$. We show in Fig~\ref{fig:zc_mass} 
the halo mass $M$ that can host a galaxy of apparent magnitude, 
$m^{Ly\alpha}_{AB}=25.3$ (of the Lyman-$\alpha$ line) at 
$z=3.1$. 
Here, $m^{Ly\alpha}_{AB}$ is the narrow band apparent magnitude and can be computed from 
$L^{Ly\alpha}_{th}$ for a fixed set of cosmological parameters.
The figure clearly shows that halos of any mass (above a minimum
mass $ M_{min}$) formed at two different redshifts can shine with the same 
magnitude at $z=3.1$. For example, when $M=10^{12} M_\odot$ (shown by dotted red 
horizontal line ) these two redshifts are approximately 3.13 and 5.9
For more details on the evaluation 
of $z_1$ and $z_2$ see Section~(3) and Fig.~(3) of J13. 

We also note that when $M<M_{min}(L^{Ly\alpha}_{th})$ the central galaxy occupation drops 
to zero, as galaxies below this mass scale can never posses a luminosity 
greater than $L^{Ly\alpha}_{th}$. For $m^{Ly\alpha}_{AB} =25.3$ and $z=3.1$ we have shown this 
threshold mass in Fig.~\ref{fig:zc_mass}.  
We have tabulated $M_{min}(L^{Ly\alpha}_{th})$ 
in Table~\ref{tab1} at each redshift. From this Table it can be seen that 
the $M_{min}$ is of the order of few times $10^{10} M_\odot$ at all redshifts 
and for all threshold magnitudes. These threshold masses of galaxies that can host an 
LAEs satisfying the threshold luminosities are roughly one order in magnitude smaller 
than lowest mass of LBGs with comparable threshold luminosities in UV light 
(see Table 3 of J13).  

\begin{figure}
\includegraphics[trim=0cm 0cm 0cm 0cm, clip=true, width =8.5cm, height=8.5cm, angle=0]
{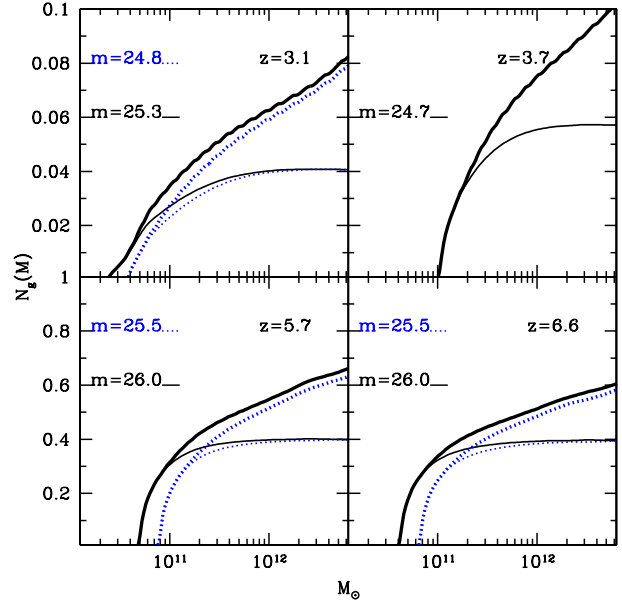} 
\caption{
The  halo occupation distribution, $\langle N_g(M)=f_{cen}(M)+N_s(M)\rangle$, as a function 
of the mass of the  hosting halo, as predicted by our models at various redshifts and threshold 
magnitudes. In each panel the thin curves corresponds to $f_{cen}(M)$ where as the thick 
curves give the total occupation $N_g(M) = f_{cen}(M)+N_s(M)$. The solid and dotted lines 
correspond to two different threshold magnitudes.
The curves are obtained using a fiducial set of model parameters (given in Section 2) that 
reproduce the observed luminosity function. In addition to that, to obtain the satellite 
occupation ($N_s(M)$), we have adopted $\Delta t_0 = 2.0 t_{dyn}$ (see text for details). 
}
\label{fig:Ng}
\end{figure}

In Fig. \ref{fig:Ng} we have plotted as thin lines, the average occupation 
number of central galaxies ($f_{cen}(M)$) as function of the mass of the 
parent halo, calculated using the above prescription at different redshifts. 
For computing the $f_{cen}$ we have used $G_f$ and $f_* f_{\rm esc}^{Ly\alpha}$ 
which best fits the UV LF and and Ly$\alpha$ LF of LAEs\footnote{Note our
LF calculations ignores the presence of satellite galaxies in the parent
dark matter halo. However J13 has shown that the inclusion of their contribution
produces very minor changes to the best fitted $f_*/\eta$.}.

\subsubsection{The satellite galaxy occupation}
Following J13 we have used the progenitor mass function and our star 
formation prescription to obtain an estimate of the number of subhalos 
and thereby the number of satellite LAEs hosted by a halo of mass $M$. 
The conditional mass function \citep{mo_white_96} gives the comoving number 
density of subhalos of mass $M_s$ which formed at $z_s$ inside a region 
containing a mass $M$ (or comoving volume $V$) that have a density contrast  
$\delta$ at $z_p<z_s$. 
To begin with, J13 consider regions of mass $M$ which have already 
formed virialized halos by the observational redshift $z$. 
%To find the number of satellite galaxies of a particular 
%luminosity inside a dark matter halo of mass $M$, J13 considers only those 
%regions of mass $M$,  is a collapsed dark matter halo at $z$. 
In this limit the conditional mass function gives the mass function of 
subhalos in the mass range $M_s$ and $M_s+dM_s$ at $z_s$ inside a halo of 
mass $M$ that collapsed at $z< z_p < z_s$. The time derivative of this conditional 
mass function is taken as the formation rate of subhalos of mass $M_s$ inside 
a big halo of mass $M$. This is because as in J13 we assume that even though 
dark 
matter halos are created and destroyed inside the over dense volume of mass $M$, 
the satellite galaxies formed in these subhalos have survived and can be 
observed. The star formation models for the satellite galaxies are taken to be the 
same as that of central galaxies (given by Eq.~(\ref{eqn:SFR})). 
This is similar to the assumption used by previous 
authors \citep{lee_09,berrier_06,conroy_wechsler_06} where 
feedback due to halo merging processes are ignored.  

We further assume in our models that no subhalo should be formed very close to the formation 
epoch of parent halo. More precisely if $t(z_p)$ is the age of the universe when a parent 
halo collapsed then all the subhalos formed inside that parent halo within a time 
interval $\Delta t_0$ prior to $t(z_p)$ do not host a satellite galaxy; rather they 
will be part of the parent halo itself. Thus in our models $t(z_p)-t(z_s) \geq \Delta t_0$ 
where $t(z_s)$ is the age of the universe when the subhalo is formed inside the parent 
halo. The time scale $\Delta t_0$ is of the order of dynamical time scale of the parent 
halo.  J13 showed that this formalism predicts the occupation of satellite galaxies  
$N_s(M, L_{th},z)$ that
can successfully explain the small scale angular clustering of 
LBGs with $\Delta t_0$ of the order of dynamical time scale $t_{dyn}$ of the parent halo. 
For predicting the occupation of Lyman-$\alpha$ emitting satellites, we again assumed 
$\Delta t_0 \geq t_{dyn}$. Interested readers may refer to Section~(4.1.2) of J13 
for more details.  

\subsubsection{The total halo occupation}
We have plotted in Fig. \ref{fig:Ng} the total occupation number of galaxies, 
$N_g(M)=f_{cen}(M)+N_s(M)$ as a function of the mass of the parent halos 
at various redshifts and for different threshold magnitudes (i.e. $m^{Ly\alpha}_{AB})$.
For each $z$ we have used the best fitted parameters given in Table~\ref{tab1}. 
These curves are shown in thick lines and thin lines correspond to 
occupation number of central galaxies. To obtain $N_s(M)$, apart from the fiducial model 
parameters that fit the observed luminosity function, we have adopted $\Delta t_0 = 2 t_{dyn}$. 
These values of $\Delta t_0$ are chosen as they are later used in Section~\ref{comparision} 
for explaining the small angular scale clustering. We find that for each limiting magnitude the 
mean numbers of central and satellite galaxies are monotonically increasing with the 
mass of the hosting halo mass. 
%As an example at $z=4$ for a halo of mass $2\times10^{12} M_\odot$ 
%the mean satellite occupation numbers are respectively 0.14, 0.27 and 0.38 
%at threshold apparent magnitudes of 25, 25.5 and 26. 
%For a bigger halo of mass $5 \times 10^{12} M_\odot$ these numbers change to 
%0.34, 0.48 and 0.61 for the same limiting magnitudes at the same redshift.

%We have also computed the average value of $N_s$ in a manner
%similar to calculating $\langle f_{cen}\rangle$ in Eq.~(\ref{fcenav}).
%The results are given in the last column of Table~\ref{tab2}.
%We see from the table that typically $\langle N_s\rangle$ is only about
%5-10\% of $\langle f_{cen}\rangle$.
%Thus the average number of detectable satellites, in a halo, 
%is typically much less
%than unity. This implies that all halos do not necessarily host an
%additional detectable satellite galaxy; however some small 
%fraction of them do
%and it is these pairs of LBGs which contribute to
%the small scale clustering. Such a conclusion has
%also been arrived at by \cite{conroy_wechsler_06}
%using numerical simulations.

It is of interest to calculate the average value of $N_g(M) = f_{cen}(M)+N_s(M)$, which is the 
average number of LAEs occupied in dark matter halos above the threshold mass $M_{min}$. 
As in J13 we define this quantity as 
\be
\langle N_g \rangle(L^{Ly\alpha}_{th},z) = 
\frac{\int_{M_{min}}^\infty dM  \left( f_{cen}(M,z) + N_s(M,z) \right) n(M,z)}
                       {\int_{M_{min}}^\infty dM n(M,z)} .
\label{fcenav}
\ee
We give the value of $\langle N_g \rangle$ in Table~\ref{tab1} for different magnitude 
thresholds and redshifts. From this table it is clear that at $z=3.1$ there is about  
1-2 \% of LAEs in dark matter halos above the threshold mass $M_{min}$. This number 
is much smaller than that of LBGs (see Table 3 of J13) which was roughly 40\%.
However as we go to higher redshift, the average $N_g$ of LAEs increases. For $z \geq 5.7$ 
this is roughly 20 \% which is roughly 50 \% of the mean occupation of LBGs at higher 
redshifts. The mean $N_g$ described above can be invoked as the duty cycle of LAEs 
(see \cite{lee_09}, J13) hence this directly implies that LAEs have a smaller duty 
cycle compared to LBGs. 

Similarly in Table~\ref{tab1}, we also give the average mass $M_{av}$ of a halo hosting 
the LAEs, for each redshift and luminosity threshold. This average mass is defined as, 
\be
M_{av} = \frac{\int_0^\infty dM  M \left( f_{cen}(M,z)+ N_s(M,z) \right) n(M,z)}
{\int_0^\infty dM f_{cen}(M,z) n(M,z)} .
\label{mav}
\ee
One can find from Table~\ref{tab1} that the average mass of LAEs of the order of 
$10^{10}-10^{11} M_\odot$ at all redshifts and threshold magnitudes we are considering. 
For example, at $z=3.1$ and for $m_{AB}^{Ly\alpha} = 25.3$ we find that 
$M_{av} = 7 \times 10^{10} M_\odot$ and at $z=5.7$ for $m_{AB}^{Ly\alpha} = 25.5$  
$M_{av} = 1.4 \times 10^{11} M_\odot$. 
Further, at a given redshift, $M_{av}$ of LAEs increases with higher threshold 
luminosity. 
We also note that, the average masses halos, 
hosting LBGs brighter than similar continuum magnitude, are higher by an order of magnitude 
(given in Table 3 of J13). 
For example at $z=4$, the average mass of LBGs 
brighter than $m_{AB}^{UV} = 25.5$ is $5.6 \times 10^{11} M_\odot$, which is 
roughly ten times bigger than that of LAEs brighter than $m_{AB}^{Ly\alpha} = 25.3$.  
This is in agreement with the suggestions that LAEs are 
residing in smaller mass halos compared to LBGs \citep{ouchi_lae_10}.

\begin{table*}
\begin{tabular}{cccccccccccc}\hline
$z$ &$f_\ast/\eta$ &$G_f$ &$f_\ast f^{Ly\alpha}_{esc}$ &$m_{AB}^{Ly\alpha}$ &$log(L_{Ly\alpha})$ 
\T\B\B\B\B &$M_{min}/M_\odot$ &$M_{av}/M_\odot$ &$\langle N_g \rangle$ &$n_g$ &$b_g$ &$b_g(O)$  \\
\hline  
\T   $\sim$ 3.1  &0.042  &0.041     &0.066  &25.3  &42.1 &$2.4\times10^{10}$ &$7.1\times10^{10}$ 
 &0.011  &1.8  &2.1  &1.5 $\pm$ 0.7  \\
\B   ~         &       &          &       &24.8  &42.3 &$3.8\times10^{10}$ &$9.6\times10^{10}$  
 &0.014  &1.1  &2.2  &1.5 $\pm$ 0.3 \\
\T\B $\sim$ 3.7  &0.038  &0.059     &0.040  &24.7  &42.6 &$10^{11}$  &$2.2\times10^{11}$  
 &0.024  &0.4  &3.5  &2.8 $\pm$ 0.5  \\
%\B  ~         &       &          &       &24.8  &42.3   &1.1  &2.2  &1.5 $\pm$ 0.3 \\
\T   $\sim$ 5.7  &0.037  &0.421     &0.031  &26.0  &42.4 &$5.0\times10^{10}$ &$9.1\times10^{10}$  
 &0.190  &1.8  &5.1  &5.5 $\pm$ 0.4  \\
\B   ~         &       &          &       &25.5  &42.6 &$7.9\times10^{10}$ &$1.4\times10^{11}$ 
 &0.180  &0.8  &5.6  &6.1 $\pm$ 0.7 \\
\T   $\sim$ 6.6  &0.044  &0.50     &0.029  &26.0  &42.4 &$4.3\times10^{10}$ &$7.5\times10^{10}$  
 &0.230  &1.1  &6.1  &3.6 $\pm$ 0.7  \\
\B   ~         &       &          &       &25.5  &42.6 &$6.8\times10^{10}$ &$1.2\times10^{11}$  
 &0.213  &0.5  &6.7  &6.0 $\pm$ 2.2 \\
\hline
\end{tabular}
\caption{
Best fitted and derived parameters of our models. 
(1) Redshift; (2) $f_\ast/\eta$, the parameter related to the light to mass ratio of galaxies; 
(3) $G_f$, the fraction of LBGs that shows detectable Ly$\alpha$ emission; 
(4) $f_\ast f^{Ly\alpha}_{esc}$, the parameter related to the net Ly$\alpha$ emission from LAEs; 
(5) the limiting magnitude of LAE samples; 
(6) approximated limiting Ly$\alpha$ luminosity corresponding to $m_{AB}^{Ly\alpha}$; 
(7) minimum mass of dark matter matter halos that can host LAEs brighter than $m_{AB}^{Ly\alpha}$; 
(8) average mass of dark matter matter halos that can host LAEs brighter than $m_{AB}^{Ly\alpha}$; 
(9) average occupation of LAEs in dark matter halos; 
(10) number density of LAEs in units of $ 10^{-3} (h/Mpc)^3$; 
(11) large scale galaxy bias of LAEs of the given threshold magnitude; 
(12) best estimate of bias by \protect \cite{ouchi_lae_10}. 
}
\label{tab1}
\end{table*}

\subsection{The correlation functions.}
\label{sec:xi_total}
In the framework of halo model, the total correlation function can be 
written as \citep{cooray_sheth_02} 
\be
\xi_g(R) = \xi^{1h}_g(R,z) + \xi^{2h}_g(R,z)
\label{eqn:xi_total}
\ee
where each term on RHS has contributions from central as well as satellite galaxies. 
On scales much bigger than the virial radius of a typical halo, the clustering amplitude 
is dominated by correlations between galaxies inside separate halos (called the 2-halo term or 
$\xi^{2h}_g$). 
On the other hand, on scales smaller than the typical virial radius of a dark matter 
halo, the major contribution to galaxy clustering is from galaxies residing in 
the same halo (called the 1-halo term or $\xi^{1h}_g$). 

The 2 halo term, $\xi^{2h}_g(r,z,L^{Ly\alpha}_{th})$, of galaxies with luminosity greater 
than $L^{Ly\alpha}_{th}$ at $z$, can now be calculated using \citep{peebles_80},
\be
\xi^{2h}_g(R,z,L^{Ly\alpha}_{th}) = \int_0^\infty \f{dk}{2\pi^2} 
k^2 \f{\sin(kR)}{kR} P^{2h}_g(k,z,L^{Ly\alpha}_{th}).     
\label{eqn:xir-2h}
\ee

Here $P_g^{2h}(k,L^{Ly\alpha}_{th},z)$ is the luminosity dependent galaxy power spectrum, 
which is computed as 
\be
P_g^{2h}(k,L^{Ly\alpha}_{th},z) = b^2_g(k,L^{Ly\alpha}_{th},z) P_{lin}(k,z).
\label{eqn:pgl}
\ee
In the above equation, $P_{lin}(k,z)$, is the linear power spectrum of CDM 
density fluctuations. The luminosity dependent galaxy bias obtained by adding 
the contributions of both the central and satellite galaxies is,
\bea
b_g(k,z,L^{Ly\alpha}_{th}) = \f{1}{n^T_g(L^{Ly\alpha}_{th},z)} \int dM n(M,z) b(M,z) \nonumber \\
       ~~~~~~~ \Big[f_{cen}(L^{Ly\alpha}_{th},M,z) +\langle N_s|L^{Ly\alpha}_{th},M,z)\rangle \Big]
	              \nonumber \\ u(k,M,z). 
\label{eqn:bias_total}
\eea 

As before, $n(M,z)$ is the ST halo mass function, and
\be
n^T_g(L^{Ly\alpha}_{th},z)= \int dM n(M,z)(f_{cen}(M,z)
+\langle N_s|L^{Ly\alpha}_{th},M,z\rangle)
\ee
%KS: shifted from below
is the total number density of galaxies, including both central and
satellite galaxies. Further, 
$b(M,z)$ is the mass dependent 
halo bias factor provided by the fitting function of Sheth and Tormen 
\citep[][also see Eq. (7) of J13]{sheth_tormen_99,cooray_sheth_02}. 
%It has the following 
%functional form 
%\be
%b(M,z) = 1+ \df{q\nu(M,z)-1}{\dl_c(z)} + \df{2p/\dl_c(z)}{1+(q\nu(M,z))^p},
%\label{eqn:halo_bias}
%\ee
%where, $\nu(M,z) = \dl_c(z)/\sigma(M)$, $\sigma(M)$ is the linearly 
%extrapolated rms density fluctuation on any mass scale $M$ 
%and $\dl_c(z)$ is the critical density required for collapse 
%at $z$. Here $\dl_c(z) = D(z) \dl_c(z=0)$ with $\dl_c(z=0) = 1.686$, 
%where $D(z)$ is the linear growth factor in a $\Lambda$CDM universe. 
%Also we use $p=0.3$ and $q=0.707$ as given by \citet{sheth_tormen_99}.
In Eq.~(\ref{eqn:bias_total}) $u(k,M)$ is the Fourier 
transform of dark matter density profile normalized by its mass, 
i.e $u(k,M,z) = \bar \rho(k,M,z)/M$. For the present 
calculations we use the Navarro Frenk White (NFW) form for the dark 
matter density distribution \citep{NFW_97}.  
On scales much larger than the virial radius of typical halos 
$u(k,M,z) = 1$ and hence the luminosity dependent galaxy bias 
given by  Eq.~\ref{eqn:bias_total} becomes a constant. 

We give in Table~\ref{tab1} this constant galaxy bias 
computed using our models. 
We note that the large scale galaxy bias 
predicted by our models compares well with observationally derived galaxy 
bias from \cite{ouchi_lae_10} (given in the last column of Table~\ref{tab1}) 
at each redshift and for nearly all threshold luminosities. 
We also find that the galaxy bias for a given $z$ increases with 
increasing luminosity and is systematically higher for high redshift 
galaxies. Further, the large scale bias, computed here for LAEs brighter than a 
given Lyman $\alpha$ threshold luminosity, is smaller than that LBGs brighter 
than the same continuum luminosity (see Table 2 of J13). 
This result has been observed previously 
by \cite{ouchi_lae_10}.  This is because, as we have seen in the last section, 
LAEs of given Lyman $\alpha$ luminosity are residing in dark matter 
halos of smaller mass compared to LBGs of the same continuum UV luminosity. 
This directly results in smaller clustering strength of LAEs compared to LBGs. 
We elaborate on this interesting fact in the next section.  

\begin{figure*}
\includegraphics[trim=0cm 0cm 0cm 0cm, clip=true, width =17cm, height=15cm, angle=0]
{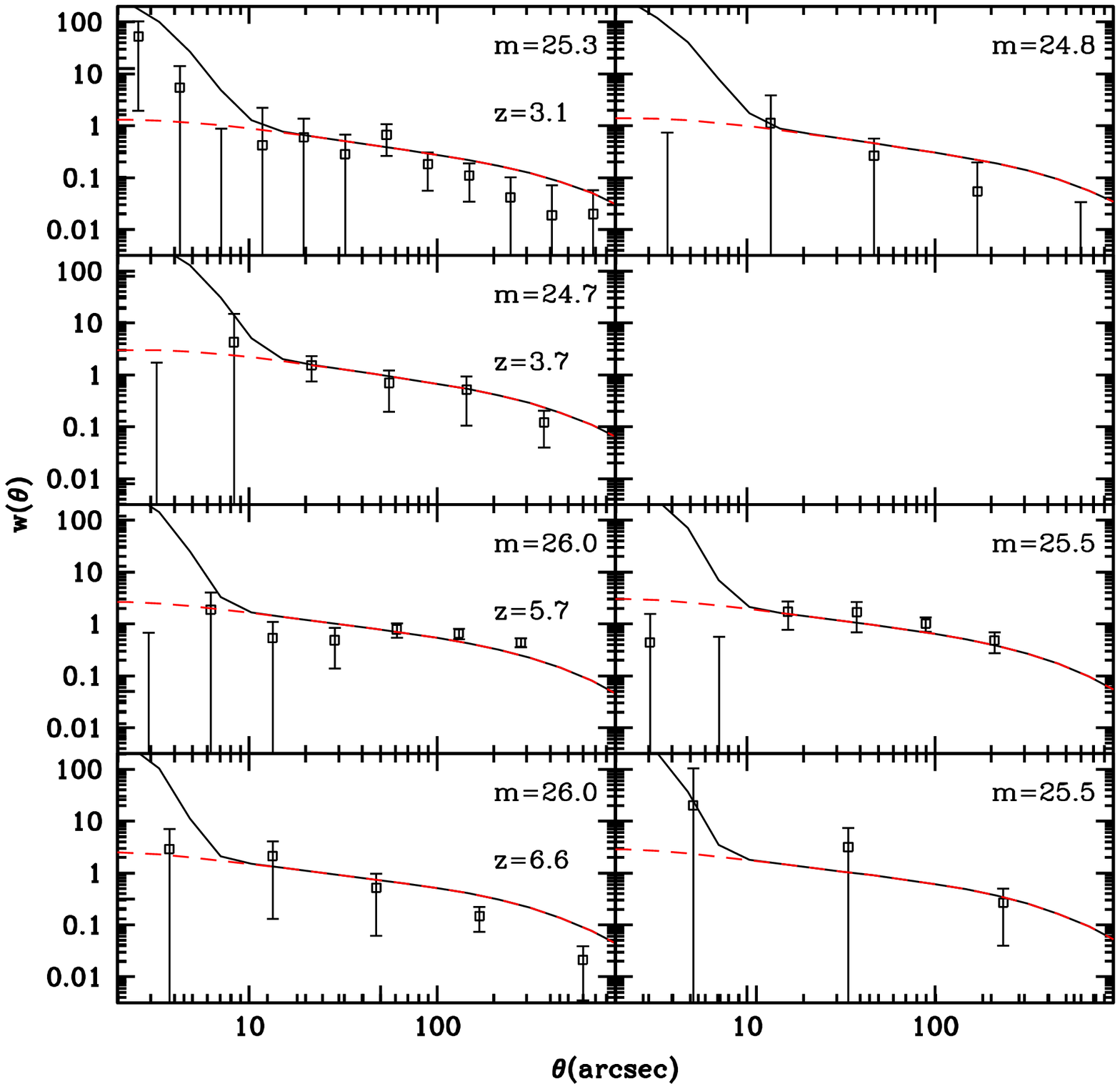} 
\caption{The angular correlation function of LAEs, taking into account 
the one halo and the two halo terms, at redshifts 3.1, 3.8, 5.7 and 6.6 
for various limiting magnitudes. Each row corresponds to a particular redshift, 
which is labelled in the first panel of that row. In each panel clustering predictions 
for galaxies with a given threshold magnitude is shown. Our fiducial model predictions of 
galaxy angular correlation functions with $\Delta t_0 = 2 t_{dyn}$ is shown in solid 
black lines. 
%The black dash-dotted 
%lines show the correlation functions computed without satellite contribution. 
The dashed red curves are our model predictions assuming a sub-Poisson distribution for 
galaxies inside dark matter halos (see text for details). 
The data points and error bars shown by black open rectangles 
are from \protect \cite{ouchi_lae_10}.
}
\label{fig:acfz_full}
\end{figure*}
\subsection{The total correlation functions}

The 1-halo term $\xi^{1h}_g(R)$ uses the standard assumption that radial 
distribution of satellite galaxies inside a parent halo follows the dark 
matter density distribution (NFW profile) \citep{cooray_sheth_02}. 
In this case the 1-halo term is given by 
\citep{tinker_weinberg_05, cooray_ouchi_06}

\bea
\xi^{1h}(L^{Ly\alpha}_{th},R,z) = \f{1}{(n_g^{T})^2}\int dM~ n(M,z)  \Big[ 2  f_{cen}(L^{Ly\alpha}_{th},M,z)  \nonumber \\ 
  ~~ \langle N_s|L^{Ly\alpha}_{th},M,z\rangle   \f{\rho_{NFW}(R|M)}{M} + \nonumber \\
  ~~ \langle N_s(N_s-1)|L^{Ly\alpha}_{th},M,z  \rangle \f{\lambda_{NFW}(R|M)}{M^2} \Big].
\label{eqn:xir-1h}
\eea
Here, as before, $n^T_g(L^{Ly\alpha}_{th},z)$
%KS: It is now already defined earlier
%\be
%n^T_g(L^{Ly\alpha}_{th},z)= \int dM n(M,z)(f_{cen}(M,z)
%+\langle N_s|L^{Ly\alpha}_{th},M,z\rangle)
%\ee 
is the total number density of galaxies which includes both the central and
satellite galaxies. Further $\rho_{NFW}$ is the NFW profile 
of dark matter density \citep{NFW_97}  inside a collapsed halo and 
$\lambda_{NFW}(r|M)$ is the convolution of this density profile with itself 
\citep{sheth_hui_01}. 
%The NFW density profile is given by 
%\be
%\rho_{NFW}(M,R) = \df{4\rho_s}{(R/R_s)(1+R/R_s)^2}
%\ee
%where the $\rho_s$ and $R_s$ are characteristic density and radius respectively.
%The ratio of the virial radius and the characteristic radius of the halo 
%is defined as the concentration parameter ($c=R_{vir}/R_s$).
For the halo concentration parameter we use the fitting functions given 
by \cite{prada_klypin_12} (Eq.~(14-23) of their paper). 
Also to begin with,
following the N-body simulations and semi-analytical 
models \citep[eg.~][]{kravtsov_04,zheng_05}, we assume that the number 
of satellites inside a parent halo forms a Poisson distribution. 
Thus we have $\langle N_s(N_s-1)\rangle =  N^2_s$, when $N_g$ is large.

Finally we compute luminosity dependent angular correlation function 
$w(\theta,z)$ from the spatial correlation function using Limber 
equation \citep[][J13]{peebles_80}
\be
w(\theta,z) = \int_0^\infty dz'~ N(z') \int_0^{\infty} dz''~N(z'') 
\xi_g\left(z,r(\theta; z',z'')\right) 
\label{eqn:limber}
\ee
where $r(\theta; z',z'')$ is the comoving separation between two points 
at $z'$ and $z''$ subtending an angle $\theta$ with respect to an observer today. 
Here we have also incorporated the normalized redshift selection function, $N(z)$, 
of the observed population of galaxies given in \cite{ouchi_lae_10,ouchi_lae_08}. 
In Eq.~(\ref{eqn:limber}) we neglect the redshift evolution of clustering of the 
galaxies detected around $z$. Hence the spatial two-point correlation function 
$\xi_g(r,z)$ is always evaluated at the observed redshift.

\section{Comparisons with observations}
\label{comparision}

The total angular correlation functions computed using our prescription for 
four redshifts and two
threshold magnitudes are overplotted on the 
observed data in Fig.~\ref{fig:acfz_full}. 
%In this figure the blue solid 
%curves are our predictions of total galaxy angular correlation functions
%for model A and the dotted curves are for model B. 
The cosmological parameters, $M_{agn}$ and $\kappa$ are kept to their fiducial 
value. All these models assumes $\Delta t_0 = 2~t_{dyn}$. 
%However at present the small scale clustering data is not good enough to test the 
%sensitivity of this parameter to clustering. 
All other model parameters are obtained by fitting the LFs and 
are given in Table~\ref{tab1}.
It is important to note that, we compare our model predictions of LF and clustering 
of LAEs with that of LAEs in the Subaru/XMM-Newton Deep Survey (SXDS) Field 
\citep{ouchi_lae_08,ouchi_lae_10,kashikawa_lae_lf_2011}. 
%In the SXDS field,  some of the galaxies detected as LAEs and thus contributing 
%to Lyman-$\alpha$ LF of LAEs are not detected as LBGs because their corresponding 
%UV continuum luminosities below the detection level. 
%However, this is not going to affect the major result, the clustering 
%predictions of LAEs, of this paper. This is because 
In particular, the clustering data of LAEs  provided by \cite{ouchi_lae_10} is for 
galaxies selected purely by their narrow band magnitude limits irrespective 
of whether those galaxies are detected in UV or not.

Firstly, we note that angular correlation functions predicted by our 
model (for $\theta \geq 10''$) is in very good agreement with the 
observed data at all redshifts and for all threshold 
luminosities. The amplitudes of these angular correlation functions are higher than 
that of LBGs (see Fig. 9 of J13) mainly because LAEs have a much 
narrower redshift distribution function $N(z)$. 
%However at $z=3.1$ and $3.7$, the large scale correlation functions 
%as predicted by model B do not compare with well with observations as compared 
%to predictions of models A.  

We  note that our models systematically overpredicts clustering at 
very small angular scales ($\theta \leq 10''$) at all redshifts and for all 
magnitudes. 
That is the contribution of one halo term predicted by our models to the $w(\theta)$ 
for small values of $\theta$ is not distinctly evident from the observed $w(\theta)$. 
We now investigate whether this discrepancy could be due to our earlier 
assumption that galaxies are Poisson distributed in dark matter halos. 
N-body simulations \citep[eg.][]{kravtsov_04} showed that 
the HOD follows a Poisson distribution when the total number of galaxies inside 
a halo exceeds unity. Number of earlier works \citep{scoccimarro_sheth_01,wechsler_01,
bullock_02} have pointed out that, galaxies follow sub-Poisson distribution 
when $N_g(M) \leq 1$. Note from Fig~\ref{fig:Ng}, the average halo occupation 
($\langle N_g \rangle$) of LAEs ($3.1 \leq z \leq 6.6$) in dark matter halos are 
typically smaller than unity. Therefore we choose the model of 
\cite{bullock_02} \citep[see also][]{hamana_04} for the average number of galaxy 
pairs inside a dark matter halo of mass $M$,
\begin{equation}
\langle N_g(N_g-1) (M) \rangle =\begin{cases}
    N^2_g(M)  & \text{$N_g\geq 1$}\\
    N^2_g(M)~\df{ln(4N_g (M))}{ln(4)} & \text{$0.25 <N_g<1$} \\
    0     &\text{$N_g\leq 0.25$}
  \end{cases}
\end{equation}
In this case the 1-halo term is computed as in \cite{hamana_04}, where 
\bea
\xi^{1h}(L^{Ly\alpha}_{th},R,z) = \f{1}{(n_g^{T})^2}\int dM~ n(M,z) \nonumber \\ 
               \langle N_g(N_g-1)|L^{Ly\alpha}_{th},M,z\rangle \f{\lambda_{NFW}(R|M)}{M^2}.
\label{eqn:xir-1h-subp}
\eea

In Fig.~\ref{fig:acfz_full} we show our model predictions (in red dashed lines) 
when one uses the above sub-Poisson distribution for galaxy occupation. 
One can see from the figure that our model predictions of angular correlation functions 
are consistent with the observed data even at very small angular scales 
($\theta \leq 10''$) for all redshifts and threshold magnitudes.
%$3.1$ and $3.7$. 
%However at $z=5.7$ our models 
%over predict small angular scale clustering. 

\begin{figure}
\begin{center}
\includegraphics[trim=0cm 0cm 0cm 0cm, clip=true, width =7.5cm, height=7.5cm, angle=0]
{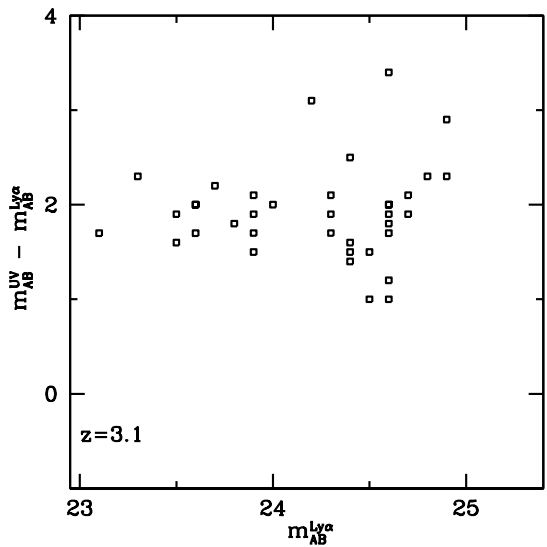} 
\caption{A comparison of the Lyman $\alpha$ AB magnitudes ($m_{AB}^{Ly\alpha}$) 
and UV AB magnitudes ($m_{AB}^{UV}$) of observed Lyman $\alpha$ emitters at $z=3.1$. 
The Y-axis shows the difference between UV and Lyman $\alpha$ magnitudes. The data points 
are from \protect \cite{ouchi_lae_08}.
}
\label{fig:obs_laes}
\end{center}
\end{figure}

As discussed earlier, galaxies selected by their threshold AB magnitude in 
Lyman $\alpha$ luminosity, are less strongly clustered than the galaxies 
selected by same UV threshold AB magnitude (See Fig. 9 and Table 2 of J13 also).   
This can be understood in the following way. In Fig.~\ref{fig:obs_laes} 
we have plotted the  difference in the observed Lyman $\alpha$ and UV 
magnitudes of LAEs at $z=3.1$ against their Lyman $\alpha$ magnitudes. 
The data points are from from \cite{ouchi_lae_08}. 
The figure clearly shows that the observed Lyman $\alpha$ apparent magnitude 
of LAEs residing in a halo of given mass are higher than their continuum 
magnitude roughly by a factor of $\Delta m_{AB} = m_{AB}^{UV} -m_{AB}^{Ly\alpha} \sim 2$. 
%Here $m_{AB}^{UV}$ and $m_{AB}^{Ly\alpha}$ are the UV and Lyman $\alpha$ 
%AB magnitudes of a given LAE. 
Or equivalently, the Lyman $\alpha$ line luminosity of LAEs are typically higher than 
their continuum UV luminosity by a factor $10^{0.8}$. 
This inturn implies that, LAEs of a given 
Lyman $\alpha$ luminosity, can be found in low mass dark matter halos compared to 
an LBG of the same UV threshold magnitude. As a result, the galaxy bias of 
LAEs of a given Lyman $\alpha$ threshold magnitude are smaller than LBGs of 
the same threshold magnitude in UV. 

The above discussions suggest that at $z=3.1$, 
the galaxy bias of LAEs with a line threshold magnitude, $m_{AB}^{Ly\alpha}$, 
must be comparable to 
the galaxy bias of LBGs with continuum threshold magnitude, $m_{AB}^{UV}$, 
where  $m_{AB}^{UV} \sim m_{AB}^{Ly\alpha} + 2$.  
In order to demonstrate this more 
elaborately, we computed the galaxy bias of LBGs (using the prescription 
of J13) at $z=3.1$  for threshold magnitude $m_{AB}^{UV} = 26.8$ 
and compared it with that of LAEs of threshold magnitude 
$m_{AB}^{Ly\alpha} = 24.8$ (here $\Delta m_{AB} = 2$). 
%We have just seen that 
%these two galaxy samples might be residing dark matter halos of similar masses, hence 
%we expect the biases of these galaxies to be the same. 
Interestingly we found that 
at $z=3.1$ galaxy bias of LBG samples (using the best fitted parameters of LBG LF) 
with $m_{AB}^{UV} = 26.8$ is $b_g= 2.35$ which compares very well with galaxy bias of 
$b_g =2.2$ that we obtain for LAEs with  $m_{AB}^{Ly\alpha} = 24.8$. 
Further, the average mass of these LBGs is $9.9 \times 10^{10} M_\odot$, 
which is very similar to the average mass, $9.6 \times 10^{10} M_\odot$, of LAEs 
with $m_{AB}^{Ly\alpha} = 24.8$.   
We performed similar exercise at other redshifts and found comparable galaxy bias and 
average mass for LAEs and LBGs when appropriate limiting magnitude are chosen.
This implies that the connection between dark matter halo and UV luminosity found by 
J13 for the LBGs is valid for the LAEs as well. 

\section{Discussion and conclusions}

We have presented here a physically motivated semi-analytical model of galaxies 
to understand the clustering of high redshift Lyman-$\alpha$ emitters.
In our models we assign luminosities (UV and Ly$\alpha$) to galaxies residing 
in dark matter halos by a physical model of star formation. 
The free parameters of this star formation model are then 
constrained by the observed high redshift luminosity functions. 
Further using the semi analytical approach of J13, we combined these 
constrained models of star formation with the
halo model and predicted 
the angular correlation functions, $w(\theta)$ of LAEs. 
Our model predictions of 
$w(\theta)$ of LAEs with different threshold magnitudes compares remarkably well 
with the observed data of \cite{ouchi_lae_10} for $\theta \geq 10''$ 
in the redshift range $z \sim 3-7$.

Our models assumed that only a fraction $G_f$ of halos hosting LBGs produce 
a detectable amound of Lyman-$\alpha$ luminosity. 
The low amplitudes of LFs LAEs compared to that of LBGs at various redshifts 
suggest that $G_f$ is typically smaller than unity. We found  
that $G_f$ ranges from $4\% - 50\%$ in the redshift range $3.1-6.6$. 
Such small values of $G_f$ directly result in a low halo occupation 
for LAEs compared to LBGs. Our models for LBG halo occupation (see J13) 
suggested that about 40\% dark matter halos above a minimum mass $M_{min}$ host 
LBGs above a threshold luminosity. On the other hand, the average number LAEs 
occupying dark matter halos, $\langle N_g \rangle$, are much smaller, ranging 
from $1\%-23\%$ in the above redshift range. 
As low halo occupation is related to duty cycle (see J13), we can conclude that 
Lyman-$\alpha$ emission has much smaller duty cycle compared to the UV emitting 
phase of the high redshift galaxies.

We find that, for observationally determined $\eta$ from \cite{bouwens_illingworth_12,
reddy_pettini_12} studies at various redshifts, the escape fraction of 
Lyman-$\alpha$ from LAEs varies from $\sim 0.5$ to $\sim 0.6$ for the redshift range 
$3.1 \leq z \leq 6.6$. Thus in our studies $f^{Ly\alpha}_{esc}$ is not strongly evolving in 
the above redshift range and such a trend is consistent with previous studies. 
The lack of sharp decrease in $f^{Ly\alpha}_{esc}$ in this redshift range is consistent 
with the epoch of reionization $z_{re} \geq 6.6$. 
Such large values of $f^{Ly\alpha}_{esc}$ are consistent with previous 
studies and can be explained by clumped models of interstellar medium 
\citep{neufeld_lae_91,dayal_lae_escape_10,forero-romero_igm_10}.

We found that the average masses, $M_{av}$ of halos hosting LAEs brighter than any 
threshold Lyman-$\alpha$ luminosity are smaller than that of LBGs  
brighter than similar continuum threshold luminosity. 
For Lyman $\alpha$ apparent magnitudes 25-26, $M_{av}$ ranges typically 
from $7.0 \times 10^{10} M_\odot - 1.2 \times 10^{11} M_\odot$ and are smaller than 
typical LBGs of similar continuum magnitude by a factor 10. Because of their smaller 
halo masses, LAEs brighter than a given threshold line luminosity have 
smaller galaxy bias and are more weakly clustered 
than LBGs brighter than similar continuum luminosities. Our model 
predictions of galaxy bias compares very well with observationally derived galaxy 
bias of \cite{ouchi_lae_10}. We also found that, the smaller masses and biases of 
LAEs brighter than a given line luminosity compared to LBGs brighter 
than similar continuum luminosity is 
the simple consequence of the fact that line luminosity of LAEs are typically higher 
than their continuum UV luminosity. In particular, Fig.~\ref{fig:obs_laes} shows that  
the difference between the observed continuum magnitude, $m^{UV}_{AB}$, and line 
magnitude, $m^{Ly\alpha}_{AB}$, of LAEs at $z=3.1$ is $\Delta m_{AB} \sim 2$. 
We further showed that the galaxy bias of LBG sample with $m^{UV}_{AB} = 26.8$ is 
almost same as the galaxy bias of LAEs with $m^{Ly\alpha}_{AB} = 24.8$. 
In addition the average mass derived for these LBGs 
($m^{UV}_{AB} = 26.8$) is very much comparable to the average mass of 
LAEs with $m_{AB}^{Ly\alpha} = 24.8$.   
This clearly suggests that LAEs belong to the same galaxy population of LBGs with 
narrow band technique having more efficiency in picking up galaxies with low 
UV luminosity or low mass. 

Finally we note that our simple model, that uses Poisson distribution for 
satellite galaxies in dark matter halos, predicts an excess in the small angular scale 
($\theta \leq 10''$) clustering. The angular correlation functions on these scales are 
dominated by one halo term. We demonstrate that this discrepancy can be removed if 
one uses a sub-Poisson distribution for galaxies inside dark matter halos  
when the total galaxy occupation $N_g$ is smaller than unity. Since $G_f$ is 
always smaller than unity, our models predict a low occupation 
of LAEs in dark matter halos (See Fig~\ref{fig:Ng}) for all redshifts 
and for various threshold magnitudes. The total occupation, $N_g$ is bigger than 
unity only for very massive dark matter halos. 
%Our models with with 
%sub-Poisson distribution for LAEs in dark matter halos consistently reproduced their 
%small scale angular clustering. 
The small angular scale clustering of LAEs can in principle 
be used constrain the statistics of galaxy occupation in dark matter halos when the mean 
occupation, $N_g$, is smaller than unity.

We have shown that the clustering of LAEs can be understood in the framework 
of halo model and physical models of galaxy formation. A clear understanding of these LAEs 
requires the complete knowledge of Lyman-$\alpha$ radiative transport in the interstellar 
medium. Nevertheless, our simple model suggests that, LAEs belong to the same galaxy population 
as LBGs and narrow band techniques are picking up galaxies with strong Lyman-$\alpha$ lines. 
 
\section*{Acknowledgments}
We thank Masami Ouchi and Nobunari Kashikawa for providing the observed data of 
luminosity functions and angular correlation functions. CJ thanks Saumyadip Samui 
for useful discussions and acknowledges support from CSIR. We thank the referee for 
valuable comments that helped us to improve the paper. 

\def\aj{AJ}%
\def\actaa{Acta Astron.}%
\def\araa{ARA\&A}%
\def\apj{ApJ}%
\def\apjl{ApJ}%
\def\apjs{ApJS}%
\def\ao{Appl.~Opt.}%
\def\apss{Ap\&SS}%
\def\aap{A\&A}% 
\def\aapr{A\&A~Rev.}%
\def\aaps{A\&AS}%
\def\azh{AZh}%
\def\baas{BAAS}%
\def\bac{Bull. astr. Inst. Czechosl.}%
\def\caa{Chinese Astron. Astrophys.}%
\def\cjaa{Chinese J. Astron. Astrophys.}%
\def\icarus{Icarus}%
\def\jcap{J. Cosmology Astropart. Phys.}%
\def\jrasc{JRASC}%
\def\mnras{MNRAS}%
\def\memras{MmRAS}%
\def\na{New A}%
\def\nar{New A Rev.}%
\def\pasa{PASA}%
\def\pra{Phys.~Rev.~A}%
\def\prb{Phys.~Rev.~B}%
\def\prc{Phys.~Rev.~C}%
\def\prd{Phys.~Rev.~D}%
\def\pre{Phys.~Rev.~E}%
\def\prl{Phys.~Rev.~Lett.}%
\def\pasp{PASP}%
\def\pasj{PASJ}%
\def\qjras{QJRAS}%2215.bib
\def\rmxaa{Rev. Mexicana Astron. Astrofis.}%
\def\skytel{S\&T}%
\def\solphys{Sol.~Phys.}%
\def\sovast{Soviet~Ast.}%
\def\ssr{Space~Sci.~Rev.}%
\def\zap{ZAp}%
\def\nat{Nature}%
\def\iaucirc{IAU~Circ.}%
\def\aplett{Astrophys.~Lett.}%
\def\apspr{Astrophys.~Space~Phys.~Res.}%
\def\bain{Bull.~Astron.~Inst.~Netherlands}%
\def\fcp{Fund.~Cosmic~Phys.}%
\def\gca{Geochim.~Cosmochim.~Acta}%
\def\grl{Geophys.~Res.~Lett.}%
\def\jcp{J.~Chem.~Phys.}%
\def\jgr{J.~Geophys.~Res.}%
\def\jqsrt{J.~Quant.~Spec.~Radiat.~Transf.}%
\def\memsai{Mem.~Soc.~Astron.~Italiana}%
\def\nphysa{Nucl.~Phys.~A}%
\def\physrep{Phys.~Rep.}%
\def\physscr{Phys.~Scr}%
\def\planss{Planet.~Space~Sci.}%
\def\procspie{Proc.~SPIED}%
\let\astap=\aap
\let\apjlett=\apjl
\let\apjsupp=\apjs
\let\applopt=\ao

\bibliographystyle{mn2e}	% (uses file "plain.bst")
\bibliography{ref.bib,sfr.bib,ref-1.bib}		% expects file "myrefs.bib"

\end{document}